\documentclass[a4paper,fleqn,usenatbib]{mnras}

\usepackage[T1]{fontenc}
\usepackage{ae,aecompl}
\usepackage{amsmath}	
\usepackage{amssymb}	
\usepackage{bm}

\newfont{\caps}{cmcsc10}

\title[CR diffusion]{Are 10 EeV cosmic rays extragalactic? Theory of cosmic ray diffusion at high energy.}

\author[A.\ Gruzinov]{
Andrei Gruzinov$^{1}$\thanks{E-mail: ag92@nyu.edu}
\\
$^{1}$Physics Department, New York University, 726 Broadway, NY
10003, USA}

\pubyear{2017}

\begin{document}
\label{firstpage}
\pagerange{\pageref{firstpage}--\pageref{lastpage}}

\maketitle

\begin{abstract}
Auger Collaboration has reported a large-scale anisotropy in the arrival directions of cosmic rays above 8 EeV. The dipole direction, at 125$^\circ$ from the Galactic center, is taken as an indication of an extragalactic origin of these cosmic rays. We show, both theoretically and by direct numerical simulations, that this is not necessarily true. Intermediate mass nuclei originating in the Galaxy and diffusing in the Galactic magnetic field can have a dipole anisotropy pointing away from the Galactic center. Our theory predicts a quadrupole anisotropy of the same order of magnitude as the dipole. 
 \end{abstract}


\section{Cosmic rays: why extragalactic?}

\label{sec:extra}

\noindent
At first sight, it would seem strange to assume that cosmic rays (CR), in any energy interval, are extragalactic. We are located at $d\approx 8$kpc from the Galactic center, the density of $L_\star$ galaxies is about $n\sim 10^{-2}{\rm Mpc}^{-3}$. For the Universe to outshine the Galaxy in CRs, even neglecting CR diffusion in the Galaxy, one would have to collect extragalactic CRs coming from the distance of about $\frac{1}{nd^2}\sim 10^3$Gpc, which is way outside the horizon. The Galaxy should dominate in CRs just as it dominates in the visible light, unless there is a large, $\sim 300$, bias in the extragalactic CR sources. 

There are only two reasons to think that some CRs are extragalactic. Consider $\sim 100$EeV CRs, and assume, just for the sake of the argument, that these are all protons. These particles travel almost undeflected by the Galactic magnetic field. If they originate in the Galaxy, the arrival directions should show anisotropy much larger than the current upper bound. 

Another argument for the etragalactic origin comes from the estimate
\begin{equation}
L\gtrsim c\left( \frac{E}{Ze} \right)^2\sim 3\times 10^{41}\times Z^{-2}E_{\rm EeV}^2\frac{\rm erg}{\rm s}
\end{equation} 
of the luminosity $L$ of a source of CRs with charge $Z$ and energy $E=E_{\rm EeV}$EeV. For $\sim 100$EeV protons, this gives $L\gtrsim 3\times 10^{45}\frac{\rm erg}{\rm s}$. We must conclude that CRs come from bursts, like gamma-ray bursts \citep{wax} or supermassive black hole flares \citep{fg}, because there are no permanent sources with $L\gtrsim 3\times 10^{45}\frac{\rm erg}{\rm s}$ within the GZK sphere. Such bright bursts are supposedly rare, and each of them, if Galactic, illuminates Earth for only about 100y. If the bursts happen in the Galaxy at a rate smaller than one per 100y, we are likely to get no Galactic CRs at all.

Both arguments in favor of extragalactic CRs disappear when we replace 100EeV protons by 8EeV intermediate mass nuclei, which we take, following \cite{ac} to have $Z=7$, representing CNO nuclei:

\begin{enumerate}

\item These CRs are strongly deflected by the Galactic magnetic field and travel through the Galaxy diffusively. They will become roughly isotropic, even if they originate in the Galaxy. 

\item The luminosity of the sources drops from $L\gtrsim 3\times 10^{45}\frac{\rm erg}{\rm s}$ to $L\gtrsim 3\times 10^{41}\frac{\rm erg}{\rm s}$, these bursts will occur more often.

\item The residence time in the Galaxy is increased from $\sim 10^4$y to about $\sim 10^6$y. This increases the density of CRs within the Galaxy.

\item The Earth illumination time is increased from $\sim 10^2$y to about $\sim 10^6$y. Now the Earth can be illuminated even if the bursts are as rare as one per $10^6$y per galaxy. 

\end{enumerate}

One does not expect perfect isotropy for CRs of Galactic origin and indeed \cite{aa} detect some anisotropy: ``The anisotropy, detected at more than the $5.2\sigma$ level of significance, can be described by a dipole with an amplitude of $6.5^{+1.3}_{-0.9}$\% towards right ascension $\alpha _d= 100\pm10$ degrees and declination $\delta_d= − 24^{+12}_{-13}$ degrees. That direction indicates an extragalactic origin for these ultra-high energy particles.''

We will show that the last statement -- ``...direction indicates an extragalactic origin...'' -- is not necessarily true. Our proposal is as follows: (i) Galactic $Z=7$, $E=8$EeV CRs have dipole anisotropy because they diffuse and their density has a gradient; (ii) on average, the dipole points along the density gradient, to the Galactic center; (iii) locally, the dipole is magnetically lensed and can occasionally point away, at more than $90^\circ$, from the Galactic center. 

We first give a qualitative discussion of CR diffusion at high energy (\S\ref{sec:qual}), then develop a rigorous theory (\S\ref{sec:theo}), and then confirm the results by a direct numerical simulation of CR propagation in magnetic field (\S\ref{sec:num}).

\section{CR diffusion at high energy: qualitative}

\label{sec:qual}

\noindent

At high energy, when the CR Larmor radius $\rho \equiv \frac{E}{ZeB}$ is much larger than the coherence length of the magnetic field $r_c$, a rigorous theory of CR diffusion can be developed. At a qualitative level, it goes as follows.

After passing one coherence length $r_c$, the CR is deflected by $\theta \sim \frac{r_c}{\rho}$. The resulting diffusion in the propagation direction is described by the angle diffusion coefficient $D_a\sim \frac{\theta ^2}{r_c/c}\sim \frac{cr_c}{\rho ^2}$. This gives the mean free path $\lambda$ -- the distance needed to deflect a CR by an angle $\sim 1$: $\lambda \sim \frac{c}{D_a}\sim \frac{\rho ^2}{r_c}$. The CR then diffuses in space with the diffusion coefficient $D\sim c\lambda$, or
\begin{equation}
D\sim \frac{c\rho ^2}{r_c}.
\label{eq:dife}
\end{equation} 
If $R$ is the ``size'' of the Galaxy, the dipole anisotropy
\begin{equation}
d\sim \frac{D}{cR}\sim \frac{\rho ^2}{r_cR}
\end{equation} 
will result. 

To see if one can get a reasonable dipole, we need to know the Galactic magnetic field, both the rms field $B$ and its coherence length $r_c$. The most detailed picture of a magnetic field in a spiral galaxy is provided by numerical simulations \citep[e.g.,][]{pak1, pak2}. Although the applicability of such numerical results is questionable, the actual mapping of the Galactic magnetic field \citep[e.g.,][]{han, far} is not in contradiction with a simulated magnetic field. As a crude summary,
\begin{equation}
B\sim 1-10\mu {\rm G},~~~r_c\sim 0.1-1{\rm kpc}
\end{equation} 
within $\sim 10$kpc radius and $\sim 3$kpc height. With $Z=7$, $E=8$EeV we get the Larmor radius
\begin{equation}
\rho \sim 0.1-1{\rm kpc}.
\end{equation} 
For the limiting case $\rho \sim r_c$, when our estimates are still roughly applicable, we get the dipole
\begin{equation}
d\sim 0.01-0.3.
\end{equation} 

On average, the dipole points to the Galactic center, under a reasonable assumption that CR sources either come from stars of from Sgr A$^*$ flares. But in the limiting case $\rho \sim r_c$, magnetic lensing rotates the local dipole by an angle $\theta \sim 1$. At some observation points, the dipole may point away rather than to the Galactic center. 

The same magnetic lensing will also give rise to higher multipole anisotropy, with a local quadrupole $Q$ of the same order of magnitude as the dipole.

In summary, assuming $\rho \sim r_c \ll R$, we expect cosmic ray anisotropy with the following characteristics:
\begin{itemize}

\item dipole anisotropy $d\sim \frac{r_c}{R}$;

\item the dipole points mostly to, but sometimes away from the Galactic canter;

\item quadrupole anisotropy $Q\sim d$.

\end{itemize}
Both theory and direct numerical simulations presented in the following sections confirm these qualitative results.

\section{Theory of CR diffusion at high energy}

\label{sec:theo}

\noindent

Rigorous theory of CR diffusion is possible for 
\begin{equation}
\rho \gg r_c.
\end{equation} 
We develop this (quite straightforward) theory below. The theory confirms the qualitative results of \S\ref{sec:qual}. 

\subsection{Magnetic field}

Consider CRs diffusing in a random static isotropic parity-invariant magnetic field ${\bf B}$. The magnetic field has zero mean 
\begin{equation}
<{\bf B}>=0,
\end{equation} 
and the second-order correlation function
\begin{equation}
<B_i({\bf r}_1)B_j({\bf r}_2)>=\Phi(r)\delta _{ij}+\Psi(r)\hat{r}_i\hat{r}_j,~~~{\bf r}={\bf r}_2-{\bf r}_1,
\end{equation} 
where, by solenoidality,
\begin{equation}
r^2\frac{d}{dr}\Phi+\frac{d}{dr}(r^2\Psi )=0,
\end{equation} 
which is solved by
\begin{equation}
\Phi =X''+\frac{1}{r}X', ~~~\Psi=-X''+\frac{1}{r}X',
\label{eq:X}
\end{equation} 
with an arbitrary $X(r)$; the reason for an extra derivative will become clear momentarily. Higher-order correlators are not needed for our purposes. 

In Fourier space,
\begin{equation}
<B_i({\bf k})B_j({\bf k}')>=(2\pi )^3\delta({\bf k}+{\bf k}')M(k)(\delta _{ij}-\hat{k}_i\hat{k}_j).
\end{equation} 
The real and Fourier space correlators are related by
\begin{equation}
X(r)=-\int \frac{d^3k}{(2\pi )^3}\frac{M(k)}{k^2}e^{i{\bf k}\cdot{\bf r}}.
\end{equation} 
Note that $M(k)$ is an arbitrary non-negative function, meaning that admissible real space correlators must be given by Eq.(\ref{eq:X}) with non-positive Fourier components of $X$.

To facilitate comparison of the forthcoming rigorous results and the qualitative results of \S\ref{sec:qual}, we must introduce the characteristic magnetic field $B$ and the coherence length $r_c$. For $B$ we just take the rms field:
\begin{equation}
B^2\equiv <B_i({\bf r})B_i({\bf r})>=3\Phi (0)=\frac{1}{\pi^2}\int_0^\infty dk~k^2M(k),
\end{equation} 
where we used $\Psi(0)=0$, as follows from Eq.(\ref{eq:X}). 

The coherence length is most naturally defined in terms of the mean wavenumber:
\begin{equation}
\frac{1}{r_c^2}\equiv {\rm const}\times \frac{\int d^3kM(k)k^2}{\int d^3kM(k)},
\end{equation} 
with some arbitrarily chosen constant. We adopt 
\begin{equation}
r_c^2\equiv -2\frac{\Phi(0)}{\Phi''(0)}=5\frac{\int_0^\infty dk~k^2M(k)}{\int_0^\infty dk~k^4M(k)}.
\end{equation} 

We will use a simple analytic correlator (with negative Fourier components of $X$): 
\begin{equation}
X=-\frac{1}{6}r_c^2B^2e^{-\frac{r^2}{2r_c^2}}
\end{equation} 
\begin{equation}
\Phi=-\left( \frac{2}{r_c^2}-\frac{r^2}{r_c^4}\right)X,~~~\Psi=-\frac{r^2}{r_c^4}X,
\end{equation} 
\begin{equation}
<B_i(0)B_j({\bf r})>=\frac{1}{6}B^2e^{-\frac{r^2}{2r_c^2}}\left( (2-\frac{r^2}{r_c^2})\delta_{ij}+\frac{r_ir_j}{r_c^2}\right),
\label{eq:model}
\end{equation} 
for concrete illustrations of the results. 

\subsection{Dipole anisotropy and diffusion coefficient}

The CR distribution function $f({\bf r},{\bf p})$ satisfies static Boltzmann equation
\begin{equation}
\hat{p}\cdot \frac{\partial f}{\partial {\bf r}}+(\hat{p}\times{\bf B})\cdot \frac{\partial f}{\partial {\bf p}}=0.
\label{eq:boltz}
\end{equation} 
Here we assume very high Lorentz factors, and units 
\begin{equation}
c=Ze=1.
\end{equation} 
We will solve the Boltzmann equation (\ref{eq:boltz}) perturbatively assuming small $B$. Here small $B$ means $\rho \gg r_c$ where 
\begin{equation}
\rho =\frac{p}{B}=\frac{E}{B},
\end{equation} 
$E$ is the CR energy. Hence, the theory works for high-energy CRs.

The unperturbed, zeroth-order, distribution function satisfies
\begin{equation}
\hat{p}\cdot \frac{\partial f^{(0)}}{\partial {\bf r}}=0.
\end{equation} 
To simplify the theory, we consider mono-energetic CR spectrum, with $E=1$. Since the actual CR spectrum is steep, and we only consider CRs above a certain threshold energy (8EeV), this simplification is admissible. We can choose the zeroth-order solution with any prescribed value of dipole anisotropy. For $d\ll 1$, we take
\begin{equation}
f^{(0)}=\delta(p-1)(1-{\bf d}\cdot {\bf p}).
\label{eq:f0}
\end{equation} 

We must say a few words about definitions. One can define the dipole ${\bf d}$, quadrupole $Q_{ij}$, etc., as moments of the distribution function
\begin{equation}
{\bf d}=<{\bf v}>_p,~~~Q_{ij}=<v_iv_j>_p-\frac{1}{3}\delta _{ij}, ...,
\label{eq:defn}
\end{equation} 
where $<...>_p$ means averaging over cosmic ray momenta at a given point in space, ${\bf v}\equiv \hat{p}$ is the CR velocity. With this definition, the maximal possible dipole, when all CRs move in the same direction, is 1 or 100\%. The definition of a dipole adopted in CR studies, including \cite{aa}, is different by a factor of $-3$. Here the minus sign is because one uses arrival directions, not the CR velocities. The factor of 3 is just a convention. To avoid terminological confusion, in Eq.(\ref{eq:f0}) and throughout the paper, we follow the convention and replace the definitions (\ref{eq:defn}) by
\begin{equation}
{\bf d}=-3<{\bf v}>_p,~~~Q_{ij}=3<v_iv_j>_p-\delta _{ij}, ...,
\end{equation}

The first-order distribution function satisfies
\begin{equation}
\hat{p}\cdot \frac{\partial f^{(1)}}{\partial {\bf r}}+(\hat{p}\times{\bf B})\cdot \frac{\partial f^{(0)}}{\partial {\bf p}}=0,
\end{equation} 
or
\begin{equation}
\hat{p}\cdot \frac{\partial f^{(1)}}{\partial {\bf r}}-\delta (p-1)({\bf p}\times{\bf B})\cdot {\bf d}=0.
\label{eq:fo}
\end{equation} 
Equation (\ref{eq:fo}) can be solved by integration over the zeroth-order unperturbed trajectory (straight line):
\begin{equation}
f^{(1)}({\bf r},{\bf p})=\delta (p-1)({\bf d}\times{\bf p})\cdot\int _{-\infty}^0d\tau {\bf B}({\bf r}+\tau {\bf p}).
\label{eq:f1s}
\end{equation} 

At second order,
\begin{equation}
\hat{p}\cdot \frac{\partial f^{(2}}{\partial {\bf r}}+(\hat{p}\times{\bf B})\cdot \frac{\partial f^{(1)}}{\partial {\bf p}}=0.
\label{eq:f2}
\end{equation} 
Now consider the large-scale part of $f$, 
\begin{equation}
F\equiv <f>,
\end{equation} 
obtained by averaging the distribution function over $B$. We have
\begin{equation}
F^{(0)}=f^{(0)},~~~F^{(1)}=0,
\end{equation} 
and Eq.(\ref{eq:f2}) gives
\begin{equation}
\hat{p}\cdot \frac{\partial F^{(2)}}{\partial {\bf r}}+<(\hat{p}\times{\bf B})\cdot \frac{\partial f^{(1)}}{\partial {\bf p}}>=0.
\label{eq:f2a}
\end{equation} 

The average in Eq.(\ref{eq:f2a}), with $f^{(1)}$ given by Eq.(\ref{eq:f1s}), is
\begin{multline}
<(\hat{p}\times{\bf B})\cdot \frac{\partial f^{(1)}}{\partial {\bf p}}>=\delta(p-1)e_{ijk}e_{nml}d_mp_j\frac{\partial}{\partial p_i} \\
\left\{ p_l\left( \int_0^\infty d\tau\Phi(\tau)\delta_{kn}+\int_0^\infty d\tau\Psi(\tau)p_kp_n \right) \right\},
\label{eq:avt}
\end{multline} 
where $e_{ijk}$ is the absolutely antisymmetric unit tensor, and the summation convention is assumed. The $\Psi$ term in Eq.(\ref{eq:avt}) is symmetric in $ln$ and drops out because of $e_{nml}$. We get the final form of the average
\begin{equation}
<(\hat{p}\times{\bf B})\cdot \frac{\partial f^{(1)}}{\partial {\bf p}}>=-\frac{1}{3D}\delta(p-1){\bf d}\cdot{\bf p},
\label{eq:avf}
\end{equation} 
where we have defined
\begin{equation}
\frac{1}{3D}\equiv 2\int_0^\infty dr\Phi(r).
\end{equation} 
With the average given by Eq.(\ref{eq:avf}), Eq.(\ref{eq:f2a}) gives
\begin{equation}
F^{(2)}=\frac{1}{3D}\delta(p-1){\bf d}\cdot{\bf r}.
\end{equation} 

To first two orders in $B$, the average distribution function is
\begin{equation}
F= \delta(p-1)\left\{ 1-{\bf d}\cdot{\bf p}+\frac{1}{3D}{\bf d}\cdot{\bf r} \right\}.
\end{equation} 
Let $n$ be the CR density given by $F$. Then the CR flux is 
\begin{equation}
{\bf j}=-\frac{1}{3}n{\bf d}=-D\nabla n,
\end{equation}  
meaning that 
\begin{equation}
D\equiv \frac{1}{6}\left( \int_0^\infty dr\Phi(r) \right)^{-1}=\frac{4\pi}{3}\left( \int_0^\infty dk~kM(k)\right)^{-1}
\label{eq:difc}
\end{equation} 
is the CR diffusion coefficient and giving a well-known relation between the dipole anisotropy and the CR density $n$:
\begin{equation}
{\bf d}=3D\frac{\nabla n}{n}.
\end{equation} 

In agreement with the qualitative result Eq.(\ref{eq:dife}), the diffusion coefficient (\ref{eq:difc}) can be written as 
\begin{equation}
D=\frac{C_D}{r_cB^2}=C_D\frac{c\rho^2}{r_c},
\end{equation} 
where $C_D$ is dimensionless number, and we assume that the magnetic field correlation function $\Phi(r)$ falls off fast enough at large $r$. For the model magnetic field Eq.(\ref{eq:model}), 
\begin{equation}
C_D=\left(\frac{2}{\pi }\right) ^{1/2}.
\end{equation} 

\subsection{Fluctuations of the dipole direction}

Now we will calculate the rms value of the dipole deflection from the mean density gradient. The zeroth-order dipole is just ${\bf d}$. The first-order dipole is 
\begin{equation}
{\bf d}^{(1)}=-3\int \frac{d^3p}{4\pi }f^{(1)}\hat{p}.
\label{eq:d1}
\end{equation} 
The rms value of the (small) angle $\chi$ between the true dipole direction and ${\bf d}$ is, to leading order,
\begin{equation}
\chi ^2=\frac{1}{d^2}<{\bf d}^{(1)2}_{\perp} >,
\label{eq:al}
\end{equation} 
where ${\bf d}^{(1)}_{\perp}$ is the first-order dipole component perpendicular to ${\bf d}$ and, as always, $<...>$ is the average over magnetic field. 

With ${\bf d}^{(1)}$ from Eq.(\ref{eq:d1}) and $f^{(1)}$ from Eq.(\ref{eq:f1s}), Eq.(\ref{eq:al}) gives, after some manipulations,
\begin{multline}
\chi ^2=9\int\frac{d^2\hat{p}}{4\pi}\int\frac{d^2\hat{q}}{4\pi}\int_0^\infty d\tau \int_0^\infty d\sigma ~ \hat{p}_\perp\cdot \hat{q}_\perp \\
\left( (\hat{p}_\perp\cdot \hat{q}_\perp)\Phi(r)+\frac{\tau \sigma}{r^2}(\hat{p}_\perp\times \hat{q}_\perp)^2\Psi(r) \right),\\
 ~~~{\bf r}\equiv \tau\hat{p}-\sigma \hat{q}.
\label{eq:alr}
\end{multline}
To simplify the integrals, we can average, at fixed  $\hat{p}$, $\hat{q}$, $\tau$, $\sigma$, over the directions of the unperturbed dipole $\hat{d}$ (with respect to which the perpendicular components $\hat{p}_\perp$ and $\hat{q}_\perp$ are defined):
\begin{equation}
<(\hat{p}_\perp\cdot \hat{q}_\perp)^2>_d=\frac{1}{15}+\frac{7}{15}(\hat{p}\cdot \hat{q})^2,
\end{equation} 
\begin{equation}
<(\hat{p}_\perp\cdot \hat{q}_\perp)(\hat{p}_\perp\times \hat{q}_\perp)^2>_d=\frac{4}{15}(\hat{p}\cdot \hat{q})\left( 1-(\hat{p}\cdot \hat{q})^2\right).
\end{equation} 
With 
\begin{equation}
\mu \equiv \hat{p}\cdot \hat{q},
\end{equation} 
we now have
\begin{multline}
\chi ^2=\frac{3}{10}\int_{-1}^{1}d\mu \int_0^\infty d\tau \int_0^\infty d\sigma ~ \\
\left( (1+7\mu^2)\Phi(r)+4\frac{\tau \sigma}{r^2}\mu(1-\mu^2)\Psi(r)\right),\\
r^2=\tau^2+\sigma^2-2\mu\tau\sigma.
\label{eq:almu}
\end{multline}
Changing the integration variable from $\mu$ to $r$, we rewrite Eq.(\ref{eq:almu}) as
\begin{multline}
\chi ^2=\frac{3}{10}\int_0^\infty d\tau \int_0^\infty d\sigma \int_{|\tau -\sigma |}^{\tau+\sigma}\frac{rdr}{\tau\sigma} ~ \\
\left( (1+7\mu^2)\Phi(r)+4\frac{\tau \sigma}{r^2}\mu(1-\mu^2)\Psi(r)\right),\\
\mu=\frac{\tau^2+\sigma^2-r^2}{2\tau \sigma}.
\end{multline}
Now change $\tau$ and $\sigma$ variables:
\begin{equation}
u\equiv \tau+\sigma,~~~v\equiv \tau-\sigma,
\end{equation} 
\begin{multline}
\chi ^2=\frac{3}{5}\int_0^\infty dr \int_r^\infty du \int_{-r}^rdv ~ \\
r\left( \frac{1+7\mu^2}{u^2-v^2}\Phi(r)+\frac{\mu(1-\mu^2)}{r^2}\Psi(r)\right),
\\
\mu=\frac{u^2+v^2-2r^2}{u^2-v^2}.
\end{multline}
After the last change of variables, $u\rightarrow ru$, $v\rightarrow rv$, we get two terms. The first one is proportional to 
\begin{equation}
\int_0^\infty dr~r\Phi(r)=0,
\end{equation}
by Eq.(\ref{eq:X}). The second term, after calculating
\begin{equation}
\int_1^\infty du \int_{-1}^1dv ~ \frac{(1-v^2)(u^2-1)(u^2+v^2-2)}{(u^2-v^2)^3}=\frac{\pi^2}{16},
\end{equation}
becomes
\begin{equation}
\chi^2=\frac{3\pi^2}{20}\int_0^\infty dr~r\Psi(r)=\frac{3}{20}\int_0^\infty dkM(k).
\label{eq:chi2}
\end{equation}

The rms dipole deflection angle is then
\begin{equation}
\chi =C_\chi Br_c=C_\chi \frac{r_c}{\rho},
\end{equation}
where $C_\chi$ is a dimensionless coefficient. For the model magnetic field Eq.(\ref{eq:model}), 
\begin{equation}
C_\chi=\frac{\pi}{2\sqrt{5}}.
\end{equation} 

Our theory works for small magnetic field, when $\rho\gg r_c$, and the dipole deflection is small. At $\rho\sim r_c$, the theory should be roughly applicable, giving $\chi \sim 1$, as has been claimed in \S\ref{sec:qual}.

\subsection{The rms quadrupole}

To first order, the quadrupole is
\begin{equation}
Q_{ij}=\int \frac{d^3p}{4\pi }f^{(1)}(3\hat{p}_i\hat{p}_j-\delta_{ij}).
\label{eq:q1}
\end{equation} 
The rms value of the quadrupole $Q$ is given by
\begin{equation}
Q^2=<Q_{ij}^2>,
\label{eq:qa}
\end{equation} 
where the average is over the magnetic field. With $Q_{ij}$ from Eq.(\ref{eq:q1}), $f^{(1)}$ from Eq.(\ref{eq:f1s}), Eq.(\ref{eq:qa}) gives, after some manipulations,
\begin{multline}
Q^2=3\int\frac{d^2\hat{p}}{4\pi}\int\frac{d^2\hat{q}}{4\pi}\int_0^\infty d\tau \int_0^\infty d\sigma ~ (3(\hat{p}\cdot \hat{q})^2-1) \\
\left( (d^2(\hat{p}\cdot \hat{q})-({\bf d}\cdot\hat{p})({\bf d}\cdot\hat{q}))\Phi(r)+\frac{\tau \sigma}{r^2}({\bf d}\cdot(\hat{p}\times \hat{q}))^2\Psi(r) \right),\\
 ~~~{\bf r}\equiv \tau\hat{p}-\sigma \hat{q}.
\end{multline}
As before, we can average over the dipole directions without changing the result:
\begin{equation}
<({\bf d}\cdot\hat{p})({\bf d}\cdot\hat{q})>_d=\frac{1}{3}d^2\hat{p}\cdot \hat{q}
\end{equation} 
\begin{equation}
<({\bf d}\cdot(\hat{p}\times \hat{q}))^2>_d=\frac{1}{3}d^2(\hat{p}\times \hat{q})^2,
\end{equation} 
define
\begin{equation}
\mu \equiv \hat{p}\cdot \hat{q},
\end{equation} 
and get
\begin{multline}
Q^2=\frac{1}{2}d^2\int_{-1}^{1}d\mu \int_0^\infty d\tau \int_0^\infty d\sigma ~ (3\mu^2-1)\\
\left( 2\mu\Phi(r)+\frac{\tau \sigma}{r^2}(1-\mu^2)\Psi(r)\right),\\
r^2=\tau^2+\sigma^2-2\mu\tau\sigma.
\end{multline}

We can now retrace the steps which lead from Eq.(\ref{eq:almu}) to Eq.(\ref{eq:chi2}) and get
\begin{equation}
Q^2=\frac{2}{3}d^2\int_0^\infty dr~r\Psi(r),
\end{equation}
or
\begin{equation}
Q=\frac{2\sqrt{10}}{3\pi}\chi d.
\end{equation}
At the limit of applicability, when $\rho \sim r_c$, and the dipole deflection angle is $\chi \sim 1$, the theory predicts a quadrupole of order the dipole, as claimed in \S\ref{sec:qual}.

\section{CR diffusion at high energy: numerical}

\label{sec:num}

\noindent

Full agreement of the qualitative analysis of \S\ref{sec:qual} and the rigorous theory of \S\ref{sec:theo} leaves little doubt that at $\rho \sim r_c$ the CRs have dipole anisotropy $d\sim \frac{\rho}{R}$, sometimes pointing away from the Galactic center, with a quadrupole of order the dipole. Still, as CR diffusion at high energy is such a simple process, one cannot resist simulating it directly, by following trajectories of charged particles in a given (random) static magnetic field.

We model the magnetic field of the Galaxy by a sum of 300 Fourier harmonics with random phases, random directions of wavevectors, random amplitudes, with ${\bf k}\cdot {\bf B}_{\bf k}=0$, with a fixed wavenumber $k$. This gives a Gaussian isotropic parity-invariant solenoidal random field with a well-defined correlation length.  We use $k^{-1}=0.1$kpc and the rms magnetic field $B=10\mu{\rm G}$.

We further specify:
\begin{itemize}

\item A CR never returns after it leaves the (magnetic) Galaxy -- a right circular cylinder with 23kpc diameter and 8kpc height. 

\item All CRs have $Z=7$, $E=8$EeV.

\item CRs are observed at $r=8$kpc and $z=0$.

\item CRs originate at $r<5$kpc and $|z|<1$kpc.

\end{itemize}

The last assumption, regarding the CR sources, is almost arbitrary. One needs to know how often Sgr A$^*$ flares with $L\gtrsim 3\times 10^{41}\frac{\rm erg}{\rm s}$ as compared to how often stars explode with a similar luminosity. In addition, one needs to know if such explosions do accelerate $Z=7$ CRs to 8EeV -- $L\gtrsim 3\times 10^{41}\frac{\rm erg}{\rm s}$ is just a necessary condition, not a sufficient one.

Different types of stars are differently distributed in the Galaxy. According to \cite{bhg} the photometric thin disk radial scale length is about 2.5kpc. On the other hand, some models for neutron star distribution put the peak density at $r=6$kpc \citep{sar}. 

Also the true magnetic field of the galaxy is certainly not an isotropic random field. The magnetic field (both observed and simulated) knows about the spiral arm structure of the Galaxy \citep[e.g.,][]{pak1, pak2, han, far}. It is also clear theoretically that the large-scale magnetic field cannot be isotropic in a differentially rotating disk.

Thus, although our numerical calculation of the CR trajectories is perfectly correct, our modeling of the magnetic field and the CR sources is poor. The results of this section have the status of an illustration.

For each of 100 randomly selected observation points (all at $r=8$kpc, $z=0$), we trace back 30,000 CR trajectories and give each trajectory a weight proportional to the time the trajectory spends in the source region ($r<5$kpc, $|z|<1$kpc) before it ultimately leaves the Galaxy. This gives us 100 simulated maps of CR arrival directions. 

For each map, we calculate the dipole ${\bf d}=-3<{\bf v}>_p$ and the quadrupole $Q_{ij}=3<v_iv_j>_p-\delta _{ij}$. We get, averaging now over the observation points,
\begin{equation}
<d^2>^{1/2}=0.26,~~~\left(<(d^2)^2>-<d^2>^2\right)^{1/4}=0.25,
\end{equation}
\begin{equation}
<Q^2>^{1/2}=0.12,~~~\left(<(Q^2)^2>-<Q^2>^2\right)^{1/4}=0.10.
\end{equation}
The dipole points away (at more than $90^\circ$) from the Galactic center in 
\begin{equation}
23\%
\end{equation}
of observation points -- just a factor of two smaller than if it were random.

\section{Summary and discussion}

Theory of cosmic ray diffusion, valid at high energy, when the Larmor radius is much larger than the magnetic field coherence length $r_c$,
\begin{equation}
\rho \equiv \frac{E}{ZeB}\gg r_c,
\end{equation}
predicts ($R$ is the ``size of the Galaxy''): 

(i) the CR diffusion coefficient
\begin{equation}
 D\sim \frac{c\rho^2}{r_c},
\end{equation}
(ii) dipole anisotropy 
\begin{equation}
d\sim \frac{\rho^2}{r_cR}\ll 1,
\end{equation}
(iii) dipole deflection angle 
\begin{equation}
\chi \sim \frac{r_c}{\rho}\ll 1,
\end{equation}
(iv) quadrupole anisotropy
\begin{equation}
Q\sim \frac{\rho}{R}\ll d.
\end{equation}

For $\rho \sim r_c$, when the theory should still be valid to order of magnitude, it predicts a dipole sometimes pointing away from the Galactic center and a quadrupole of order the dipole. 

Our results do not mean that all or some 10EeV cosmic rays are Galactic, but they do mean that all or some 10EeV cosmic rays {\it can be} Galactic. In particular, the observed dipole anisotropy {\it can be} of a purely Galactic origin, even though the dipole points away from the Galactic center. 

An illustrative calculation of \S\ref{sec:num} gives the rms dipole equal to 26\%. The observed dipole is 6.5\%. One can then propose the following interpretation. Most of the observed 10EeV cosmic rays are extragalactic, and these are isotropically distributed in the sky. But some 20\% of the observed 10EeV cosmic rays originated in the Galaxy. These CRs are highly anisotropic. It just so happens, that the (inevitable) dipole anisotropy of the Galactic CRs is magnetically lensed away from the Galactic center.

A quadrupole measurement might become possible with better CR statistics. Our theory firmly predicts a quadrupole of order the dipole (a factor of 2 smaller in the illustrative numerical calculation).

\section*{Acknowledgements}
I thank Glennys Farrar, Michael Unger and Eli Waxman for many CR discussions. I thank Ann Zabludoff and Dennis Zaritsky for help with stars in the Galaxy.

\label{lastpage}
\end{document}